\documentclass[a4paper,fleqn,usenatbib]{mnras}

\usepackage{newtxtext,newtxmath}
\usepackage[T1]{fontenc}
\usepackage{ae,aecompl}

\usepackage{graphicx}	% Including figure files
\usepackage{amsmath}	% Advanced maths commands
\usepackage{amssymb}	% Extra maths symbols

\title[LP~358-499]{Three small transiting planets around the M dwarf host star LP~358-499}

\author[R. Wells, K. Poppenhaeger and C. A. Watson]{
R. Wells$^{1}$\thanks{E-mail: \href{mailto:rwells02@qub.ac.uk}{rwells02@qub.ac.uk}},
K. Poppenhaeger$^{1,2}$ and
C. A. Watson$^{1}$
\\
$^{1}$Astrophysics Research Centre, Queen's University Belfast, Belfast BT7 1NN, UK
\\
$^{2}$Harvard-Smithsonian Center for Astrophysics, 60 Garden Street, Cambridge, MA 02138, USA
}
\date{Accepted XXX. Received YYY; submitted in original form 06. Sept. 2017}

\pubyear{2017}

\begin{document}
\label{firstpage}
\pagerange{\pageref{firstpage}--\pageref{lastpage}}
\maketitle

\begin{abstract}
We report on the detection of three transiting small planets around the low-mass star LP~358-499 (K2-133), using photometric data from the Kepler-K2 mission. Using multiband photometry, we determine the host star to be an early M dwarf with an age likely older than a Gigayear. The three detected planets K2-133 b, c, and d have orbital periods of ca.\ 3, 4.9 and 11 days and transit depths of ca.\ 700, 1000 and 2000 ppm, respectively. We also report a planetary candidate in the system (EPIC~247887989.01) with a period of 26.6 days and a depth of ca.\ 1000 ppm, which may be at the inner edge of the stellar habitable zone, depending on the specific host star properties. Using the transit parameters and the stellar properties, we estimate that the innermost planet may be rocky. The system is suited for follow-up observations to measure planetary masses and JWST transmission spectra of planetary atmospheres.
\end{abstract}


%%%%%%%%%%%%%%%%% BODY OF PAPER %%%%%%%%%%%%%%%%%%

\section{Introduction}

Small stars provide a very favourable opportunity to study the properties of planets orbiting around them. A planet transiting a small star will cause a deeper transit signature than in a system with a larger, more massive host star; equally, with a low-mass host star, a planet causes a stronger radial velocity signature in the stellar spectrum. Furthermore, habitable zones around the host star, i.e.\ orbital distances at which water in liquid form could be present on planets, are located relatively close to the star, making discoveries of temperate planets easier. Two notable examples of recent planet discoveries around M dwarfs are Proxima Centauri~b \citep{Anglada2016} and the multi-planet system around Trappist-1 \citep{Gillon2017}. 

\begin{figure}
\begin{center} 
\includegraphics[width=0.49\textwidth]{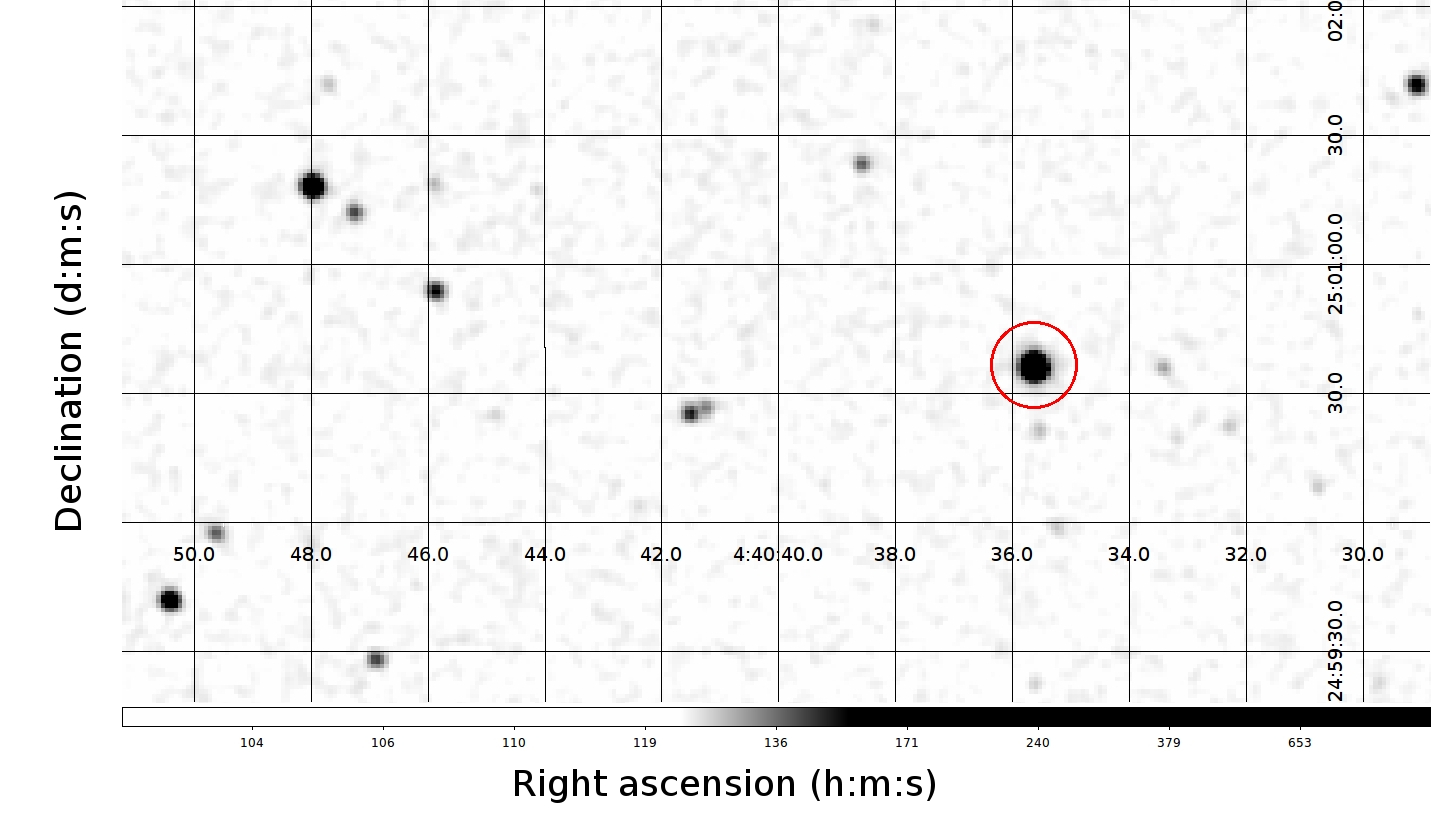} 
\caption{2MASS image of the star LP~358-499 and its vicinity; the red circle around LP~358-499 has a radius of 10~arcseconds. No potential blends are visible in the image.\label{fig:2MASS}} 
\end{center} 
\end{figure}

Here we present the discovery of three transiting planets and one planet candidate around the low-mass star LP~358-499, also known as 2MASS J04403562+2500361, NLTT 13719, EPIC~247887989, and K2-133. The star is located near the ecliptic plane; its basic astrometric properties are listed in Table~\ref{tab:sed}. The star is not located in a crowded field; there are no other bright sources visible within a $10^{\prime\prime}$ radius in either {\it Gaia} data release 1 \citep{2016A&A...595A...2G}, or in 2MASS data as shown in Figure~\ref{fig:2MASS}. We describe our data analysis methods in section~\ref{methods}, our results in section~\ref{results}, and discuss our findings regarding the nature of the planets and the stability of the system in section~\ref{discussion}.

\section{Methods}\label{methods}
\begin{figure}
\begin{center} 
\includegraphics[width=0.49\textwidth]{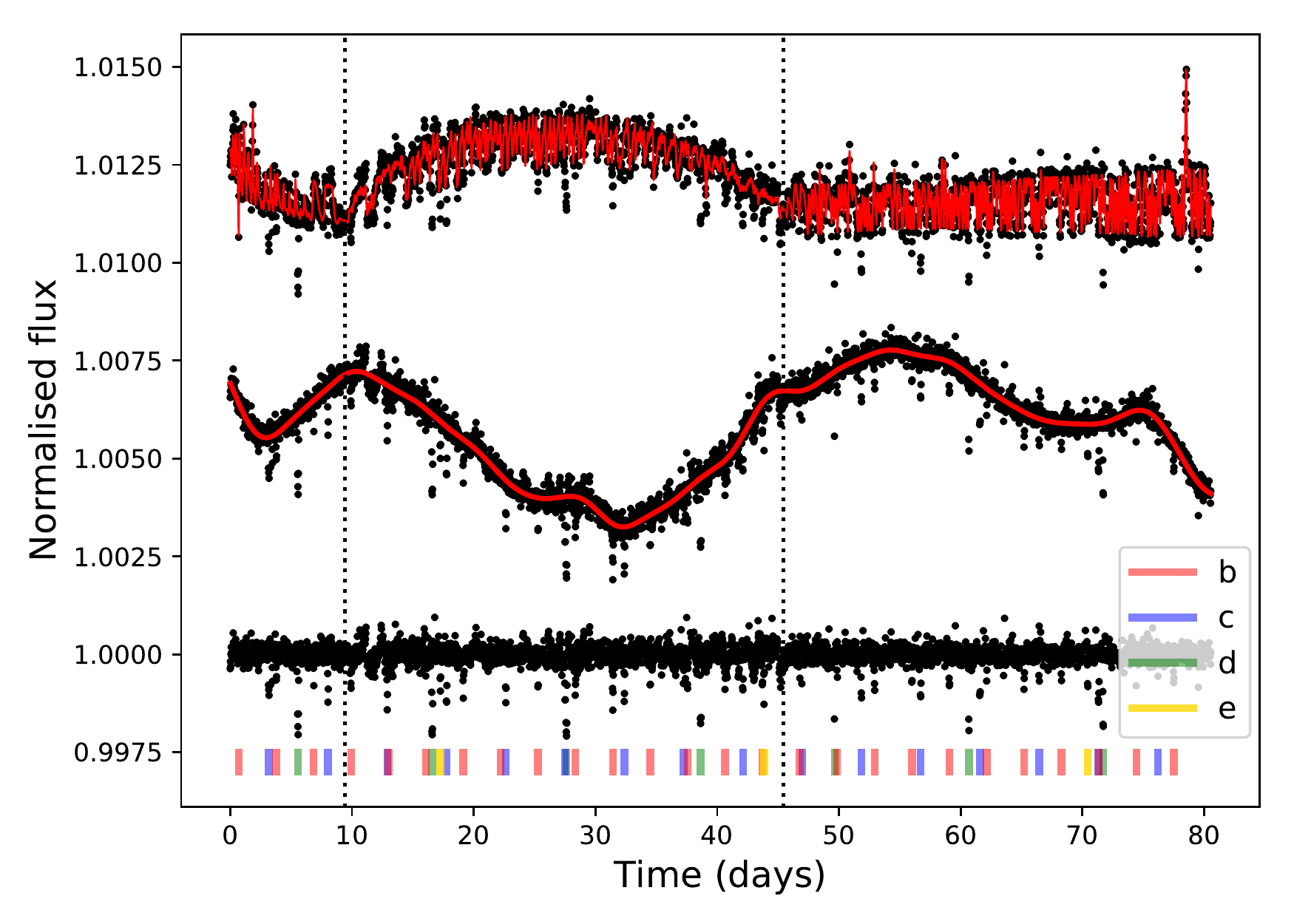} 
\caption{Systematics in the light curve of LP~358-499 observed in campaign 13. From top to bottom: flux corrected for the time-dependent trend, flux corrected for position-dependent trend and flux with both trends removed. The data are shown as black points and the models are shown in red. The vertical dotted lines show the points where the direction of the roll-angle variations reverse. The coloured markers are placed at times of the transits of each planet. \label{fig:detrend}} 
\end{center} 
\end{figure}

The target LP~358-499 was observed during Campaign 13 of the K2 mission \citep{howell2014} for 80 days, between March 08 and May 27 2017. Pre-search Data Conditioning (PDC) light curves were obtained on August 28 2017 when they became available from the Kepler team. PDC light curves have been corrected for common trends between many stars, although these still include stellar variability and detector-position systematics. To remove these effects, but leave the transit signals untouched, we detrended the light curve using the {\sc k2sc} code \citep{Aigrain2016} which models the flux as a Gaussian process formed of three components. The first component depends on the star's two-dimensional ($x, y$) position on the detector, which changes due to the radiation pressure from the Sun causing a slight drift of the telescope which is periodically corrected by thruster firings. We use values of 9.42 and 45.42 days from the campaign start as the points where the direction of these variations reverse. The second part depends on the time of the measurement and represents the stars variability plus any time-dependant systematics. The final term consists solely of white noise. Fig.~\ref{fig:detrend} illustrates the algorithm for the target of this work. The photometric precision was 1082\,ppm before detrending and 286\,ppm after; an improvement of approximately 3.8 times. As we will show in section~\ref{results}, the four transit signals identified have depths which range from ca.\ 700 to 2000 ppm.

\section{Results}\label{results}

\subsection{Properties of the host star LP~358-499}\label{host-star}

The star LP~358-499 has been observed in several optical and infrared bandpasses (see Table~\ref{tab:sed}), which we use to derive the basic physical properties of the star.

\begin{figure}
	\begin{center} 
		\includegraphics[width=0.49\textwidth]{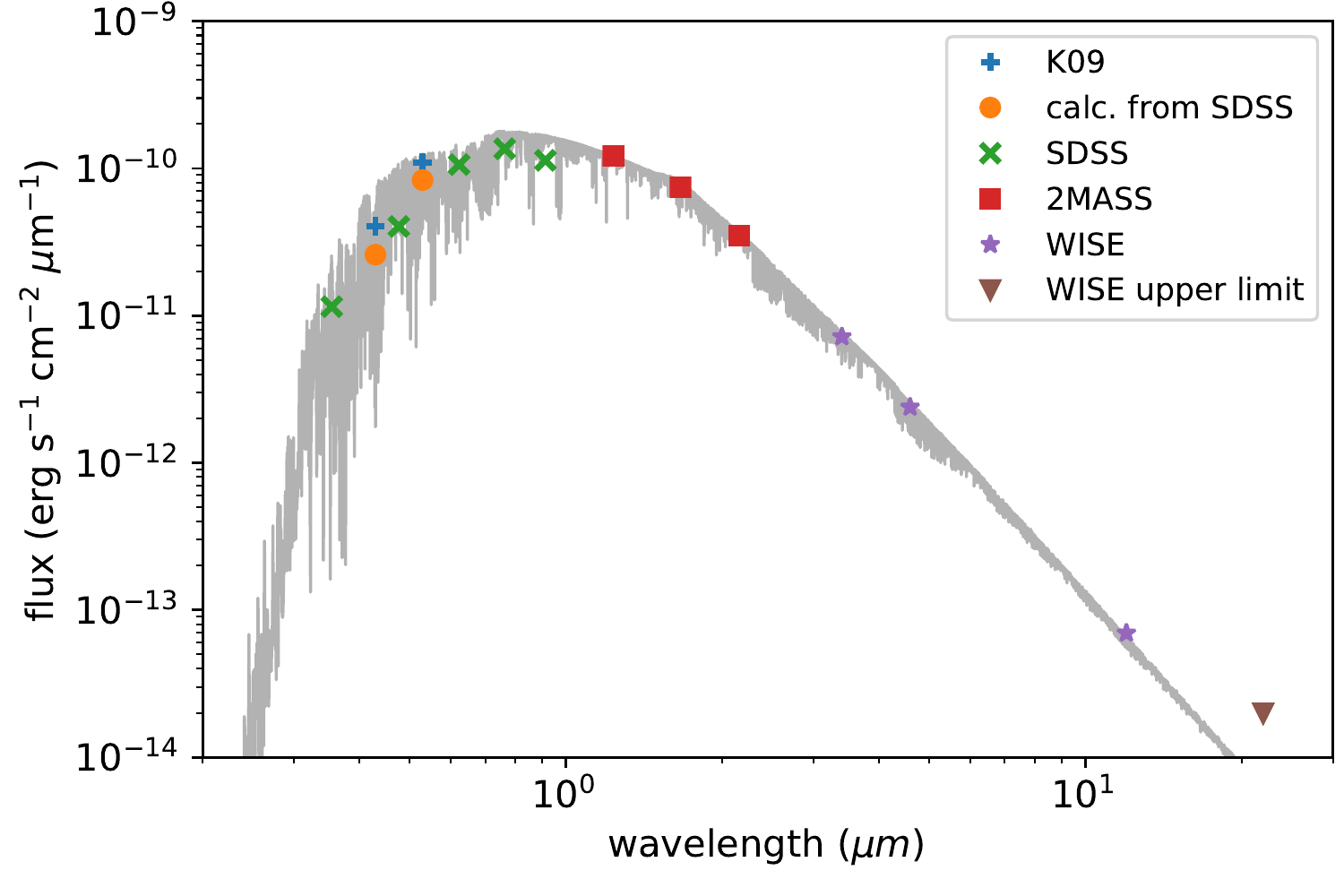} 
		\caption{Spectral energy distribution of LP~358-499, using the photometric measurements from Table~\ref{tab:sed} (see text for details). Also depicted is a theoretical SED model from the BT-Settl-CIFIST grid (grey) for an effective temperature of 3700\,K and $\log g =5.0$, scaled according to our estimated stellar radius and distance.  \label{fig:sed}} 
	\end{center} 
\end{figure}

We begin by showing the spectral energy distribution (SED) of the star in Figure~\ref{fig:sed}. No infrared excess is observed. The values for the $V$ and $B$ band, as reported by \citet{Kharchenko2009} (K09 from here on), have significantly larger uncertainties than the other photometric measurements, and seem systematically brighter than expected from the other data points in the SED. For comparison, we used the transformations from \citet{Jester2005} to calculate the expected $V$ and $B$ band magnitudes from the available high-precision SDSS $ugriz$ photometry. This yields calculated values of $V_\mathrm{calc} = 14.288 \pm 0.010$ and $B_\mathrm{calc} = 16.114 \pm 0.031$. We display those calculated values in the SED as well; they seem to be more in line with the remaining data points in the SED. We will use the SDSS-calculated $V_\mathrm{calc}$ and $B_\mathrm{calc}$ magnitudes as the preferred values in the remainder of our analysis, but will give results using the K09 magnitudes for comparison as well.

To estimate the stellar parameters we use the empirical relationships by \citet{Mann2015}. The stellar effective temperature can be calculated (for targets where the metallicity is not known a priori) from the empirical relationship $T_{\mathrm{eff}} = a + bX + cX^2 + dX^3 + eX^4 + f(J-H) + g(J-H)^2$, with $X$ being the $V-J$ colour, and the coefficients $a$ through $g$ being $2.769$, $-1.421$, $0.4284$, $-0.06133$, $0.003310$, $0.1333$, and $0.05416$, respectively -- note that an erratum has been published on these values, see \citet{Mann2016}. Using the 2MASS $J$ and $H$ magnitudes, and our SDSS-calculated $V_\mathrm{calc}$ magnitude, we find an effective temperature of $T_{\mathrm{eff}} =3655\pm 80$\,K. For comparison, using the K09 $V$ magnitude and its uncertainty, the effective temperature estimate changes to $T_{\mathrm{eff}} =3802\pm 116$\,K. 

We verified this empirical estimate by fitting the SED based on the photometric measurements listed in Table~\ref{tab:sed} with the VOSA SED fitting tool \citep{Bayo2008}. We fitted a BT-Settl-CIFIST model \citep{Baraffe2015} to LP~358-499's fluxes, finding a best fit at an effective temperature of $T_{\mathrm{eff}} =3700$\,K and a surface gravity of $\log g = 5.5$; a model with the same effective temperature and a surface gravity of $\log g = 5.0$ produces a similarly good fit.

In the following, we adopt the effective temperature of $T_{\mathrm{eff}} =3655\pm 80$\,K, found through the empirical relationship by \citet{Mann2015}, as our estimate for LP~358-499. Using the tabulated stellar properties by \citet{Pecaut2013}, we estimate LP~358-499's stellar mass as $0.51\pm 0.03\,M_\odot$, its absolute $J$ band magnitude as $M_J = 6.54\pm 0.2$, and its bolometric luminosity as $L_\mathrm{bol} = (1.4\pm 0.3)\times 10^{32}$~erg/s, which is 3.6\% of the solar bolometric luminosity. This corresponds to a spectral type of M1V, and implies a stellar radius of $0.47\pm 0.03\,R_\odot$.

Comparing the absolute and apparent $J$ band magnitudes places LP~358-499 at a distance of $78\pm 7$\,pc. The proper motion of LP~358-499 is rather large with $\mu_{\mathrm{RA}} = 187 \mathrm{mas\,yr^{-1}}$ and $\mu_{\mathrm{Dec}} = -45 \mathrm{mas\,yr^{-1}}$, i.e.\ a total proper motion of $\mu = 195 \mathrm{mas\,yr^{-1}}$. Given our distance estimate, this corresponds to a tangential space velocity of ca. 74\,km/s. This is somewhat fast for ordinary stars in the solar neighbourhood, but not completely uncommon. For comparison, Barnard's star has a tangential velocity of ca.\ 90\,km/s.

We also note here that the proper motion of the star is compatible with the star being part of the old disk or galactic halo population; this might mean it is an old star with low metallicity. However, caveats apply as the identification of a star being part of the old disk or halo population based on proper motions can be ambiguous \citep{Gizis1997}. For completeness, repeating the calculation of the effective temperature with an ad-hoc assumption of low metallicity [Fe/H]=-1 and using eq.\ (7) of \citet{Mann2015}, we find a lower effective temperature of $3503\pm 79$\,K; the stellar mass, radius, bolometric luminosity, and distance then follow as $0.46\pm 0.02\,M_\odot$, $0.41\pm 0.03\,R_\odot$, $(0.9\pm 0.2)\times 10^{32}$~erg/s, and $63\pm 7$\,pc. If the host star was indeed this metal-poor, it would affect our estimate of the planetary parameters. We therefore give planetary parameters for the low-metallicity options in brackets in Table~\ref{tab:planetparams}. Given the difference in the estimated distance, an accurate parallax, which will become available with the second {\it Gaia} data release, will significantly improve the estimate of the host star properties.

\begin{table}
\caption{Stellar properties of LP~358-499. Derived stellar parameters are given for a solar metallicity star, and alternatively for a low-metallicity star in brackets. See text for details.}
\begin{tabular}{l r r l}
\hline \hline
Property & Value & Source \\ \hline
%& & \\
Astrometry: & & \\
R.A. & 04 40 35.63 & SIMBAD\\
Dec & +25 00 36.1 & SIMBAD\\
$\mu_{\mathrm{RA}}$ (mas\,yr$^{-1}$)& 187 & SIMBAD\\
$\mu_{\mathrm{Dec}}$ (mas\,yr$^{-1}$)& -54 & SIMBAD\\[0.1cm]
%& & \\
Photometry: & & \\
$B$ (mag) & 15.633 $\pm$ 0.204 & K09\\
$V$ (mag) & 13.996 $\pm$ 0.151 & K09\\
$u$ (mag) & 17.283 $\pm$ 0.01 & SDSS \\
$g$ (mag) & 15.265 $\pm$ 0.004 & SDSS \\
$r$ (mag) & 13.626 $\pm$ 0.003 & SDSS \\
$i$ (mag) & 12.915 $\pm$ 0.001 & SDSS \\
$z$ (mag) & 12.719 $\pm$ 0.005 & SDSS \\
$B_\mathrm{calc}$ (mag) & 16.114 $\pm$ 0.031 & calc.\\
$V_\mathrm{calc}$ (mag) & 14.288 $\pm$ 0.010 & calc.\\
$J$ (mag) & 11.084 $\pm$ 0.021 & 2MASS\\
$H$ (mag) & 10.487 $\pm$ 0.021 & 2MASS\\
$K$ (mag) & 10.279 $\pm$ 0.018 & 2MASS\\
$W1$ (mag) & 10.173 $\pm$ 0.022 & WISE\\
$W2$ (mag) & 10.072 $\pm$ 0.021 & WISE\\
$W3$ (mag) & 9.991 $\pm$ 0.071 & WISE\\
$W4$ (mag) & $>$8.586  & WISE\\[0.1cm]
%& & \\
Derived properties: & & \\
$T_{\mathrm{eff}}$ & \multicolumn{2}{l}{$3644\pm 80$ ($3505\pm79$)\,K} \\
stellar radius & \multicolumn{2}{l}{$0.47 \pm 0.03$ ($0.41 \pm 0.03$) $R_\odot$} \\
stellar mass & \multicolumn{2}{l}{$0.51 \pm 0.03$ ($0.46 \pm 0.02$) $M_\odot$} \\
stellar luminosity & \multicolumn{2}{l}{$0.036 \pm 0.007$ ($0.023 \pm 0.005$) $L_\odot$} \\
distance & \multicolumn{2}{l}{$78 \pm 7$ ($63 \pm 7$) pc} \\
\hline
\end{tabular}
\label{tab:sed}
\end{table}

The star LP~358-499  has not been observed with modern X-ray telescopes (\textit{XMM-Newton} and \textit{Chandra}). Its position has been observed in the \textit{ROSAT} All-Sky Survey for 450 seconds, but the star was not detected in X-rays. \textit{ROSAT} places an upper limit of $F_\mathrm{X} < 3.0\times 10^{-13}$~ergs/s/cm$^2$ on its X-ray flux and a limit of $L_\mathrm{X} < 2.2\times 10^{29}$~ergs/s on its X-ray luminosity. This is fairly non-restrictive for an early-type M dwarf; M and K stars at the age of the Pleiades (100~Myr) have already decreased their X-ray luminosities enough to display an average $L_X$ around $10^{29}$~ergs/s, see \citet{Preibisch2005}. We can conclude that the system is not extremely young, as M dwarf in the Orion Nebula Cluster with an age of 2.5~Myr are typically X-ray brighter than the upper limit for our target star. Furthermore, from a visual inspection of the de-trended light curve in Fig.~\ref{fig:detrend} (middle light curve), a long-term modulation of ca.\ 50 days is visible, which is likely the stellar rotation period. This means that LP~358-499 is likely an older star -- for an M dwarf, such a rotation period would be typical for an age older than a gigayear \citep{Barnes2003}.

\subsection{Properties of the planet candidates}

We searched for transit signals in the detrended light curve with periods longer than 1 day using the PyBLS\footnote{\url{https://github.com/hpparvi/PyBLS}} {\sc python} package for the Box Least Squares (BLS) algorithm of \citet{KovacsBLS}, smoothed with a median filter. We iteratively identified signals with a Signal Detection Efficiency (SDE) greater than 8, as recommended by \citet{Aigrain2016} from injection tests. We then fitted the light curve with a transit model and subtracted the transit model from the light curve to search for further planet candidates. 

\begin{figure}
\begin{center} 
\includegraphics[width=0.49\textwidth]{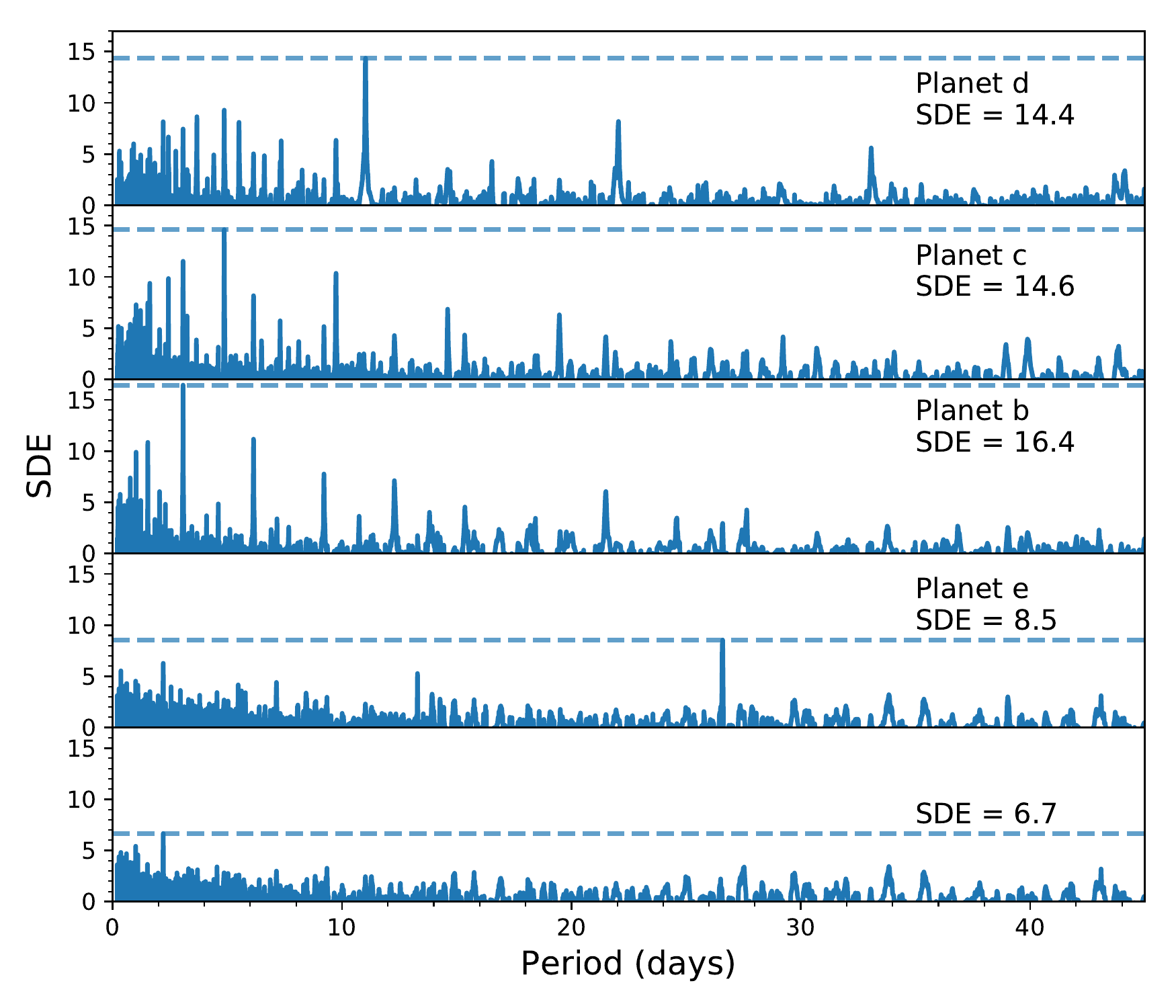} 
\caption{BLS spectra of the planets found, in the order which they were identified. The top panel shows the BLS spectrum for the detrended light curve and subsequent panels show the BLS spectra where the transit signals of the previously identified planets have been removed. The final panel shows the BLS spectrum of the light curve with all four planets removed, where no further significant periodic signals are found. \label{fig:bls}} 
\end{center} 
\end{figure}

We display the results of the BLS algorithm applied to LP~358-499's Kepler-K2 light curves in Figure~\ref{fig:bls}. We identify four transiting exoplanet candidates with orbital periods of ca.\ 3, 4.9, 11 and 26.6 days, having 26, 16, 7 and 3 transit events covered in the light curve, respectively. We note the longest period planet is not as significant as the other three due to having observed only three transits; its SDE is only slightly above the detection threshold of 8. We therefore do not treat it as a verified planet in this work yet, and recommend follow-up observations. The peak of the fifth iteration is below our detection threshold for a candidate and no transit-shaped feature was evident when phase-folding at this period.

The resulting planetary orbital periods yielded by the BLS algorithm were then used in a more detailed light curve fit, using the analytic transit light curve models by \citet{Mandel2002} as implemented in the fitting package of PyAstronomy\footnote{\url{https://github.com/sczesla/PyAstronomy}}. We performed MCMC fits of the analytic light curve models to the data, varying the orbital period, transit mid-point $T_0$, ratio of planetary to stellar radius $R_p/R_s$, and orbital inclination. The semi-major axis was linked to the orbital period through the stellar parameters derived above. We used a quadratic limb-darkening law with coefficients fixed at 0.5079 and 0.2239 as given by \citet{sing2010}; we tested varying the limb-darkening coefficients, but found that our fits were not very sensitive to them.

We list all derived properties of the three planets in Table~\ref{tab:planetparams}, along with values for the tentative planet e. We note that while the transit depth for the candidate is similar to planet c, the radius is larger due to the best fit model having an impact parameter of 0.93, and the transit having a slight V-shape compatible with a grazing transit.

\begin{figure*}
\begin{center} 
\includegraphics[width=\textwidth]{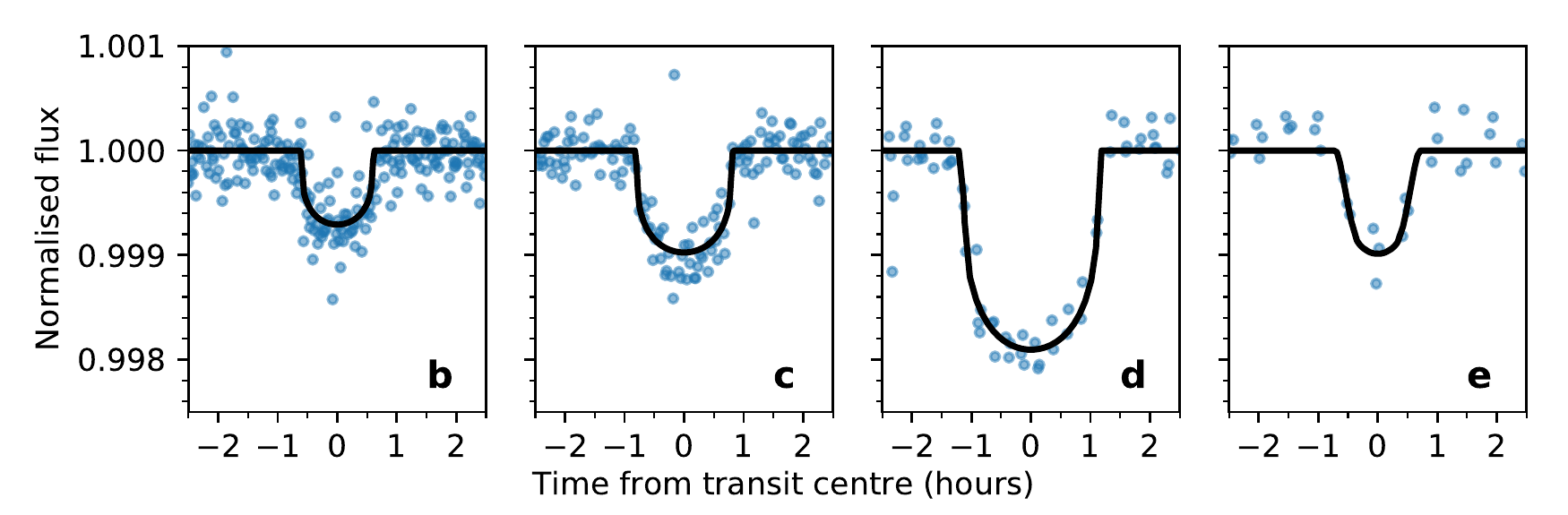} 
\caption{Phase-folded detrended transit light curves for each planet. K2 data points are shown in blue with the best fitting transit model overplotted as black lines. The the planetary parameters of these fits are given in Table~\ref{tab:planetparams}. \label{fig:phase_lcs}} 
\end{center} 
\end{figure*}

Phase-folded light curves for each planet are given in Fig.~\ref{fig:phase_lcs}, where the best fit limb darkened transit model from \citet{Mandel2002} is overplotted using the PyAstronomy {\sc python} package.

\begin{table*}
%\centering
\caption{Planetary parameters of the candidates with uncertainties are given at the 95\% confidence level. Values in brackets are for a metal-poor host star, where these are only given if values deviate by more than the related uncertainty.}
\resizebox{\textwidth}{!}{
\begin{tabular}{lcccc}
\hline\hline
Property & K2-133 b & K2-133 c & K2-133 d & (EPIC~247887989.01)\footnotemark \\
\hline
Fitted parameters: & & \\ [0.1cm]
Period (days) & $3.0712 \pm 0.0001$ & $4.8682_{-0.0003}^{+0.0001}$ & $11.0234_{-0.0003}^{+0.0008}$ & $26.5837_{-0.0020}^{+0.0033}$ \\ [0.1cm]
$T_{0}$ (BJD) & $2457821.3168_{-0.0014}^{+0.0011}$ & $2457823.7656_{-0.0009}^{+0.0023}$ & $2457826.1739_{-0.0013}^{+0.0007}$ & $2457837.8659_{-0.0033}^{+0.0029}$ \\ [0.1cm]
Inclination ($^\circ$) & $87.56_{-0.05}^{+0.06}$ $\left(88.20 \pm 0.08\right)$ & $88.65_{-0.16}^{+0.08}$ $\left(89.20_{-0.13}^{+0.39}\right)$ & $89.52_{-0.06}^{+0.15}$ $\left(89.99_{-0.25}^{+0.01}\right)$ & $89.16_{-0.01}^{+0.03}$ $\left(89.25_{-0.01}^{+0.04}\right)$ \\ [0.1cm]
$R_{p}/R_{s}$ & $0.0255_{-0.0008}^{+0.0007}$ & $0.0288_{-0.0006}^{+0.0008}$ & $0.0393_{-0.0007}^{+0.0006}$ & $0.0372_{-0.0055}^{+0.0029}$ \\ [0.1cm]
Duration (hours) & $1.22_{-0.02}^{+0.04}$ & $1.62_{-0.06}^{+0.04}$ & $2.32_{-0.09}^{+0.08}$ & $1.35_{-0.09}^{+0.19}$ \\ [0.1cm]
Depth at mid-transit (ppm) & $710 \pm 40$ & $980_{-40}^{+50}$ & $1900 \pm 60$ & $990_{-200}^{+130}$ \\ [0.5cm]
Derived parameters: & & \\ [0.1cm]
Radius ($R_{\earth}$) & 1.31 $(1.11) \pm 0.08$ & 1.48 $(1.27) \pm 0.09$ & 2.02 $(1.76) \pm 0.13$ & 1.91 $(1.63) \pm 0.12$ \\ [0.1cm]
a (au) & $0.033 \pm 0.002$ & $0.045 \pm 0.003$ & $0.077 \pm 0.005$ & $0.139 \pm 0.009$ \\ [0.1cm]
\hline
\end{tabular}}
\label{tab:planetparams}
\end{table*}

\section{Discussion}\label{discussion}

\subsection{Planetary masses}\label{sec:masses}
Radial velocities are often used to rule out false positives for planetary transits \citep{2000ApJ...529L..41H,2004ApJ...613L.153A}; those might be low-mass stars in a grazing transiting orbit, or eclipsing binaries in the background which are blended with the target star in the telescope's point spread function. Here, we do not have radial velocity measurements of the target system as of yet. 

However, we detect four different transit signatures around the star LP~358-499. It is extremely unlikely that a non-planetary, false-positive system configuration could produce those four different transit signatures. \citet{fressin} found the false positive rate for small planets found in the Kepler mission to be ca.\ 10\%. The probability to have multiple false positives is much lower; \citet{lissauer} show the probability of a system with 3 or more planet candidates having 1 or 2 false positives are 0.2\% and 0.0001\% respectively, assuming a 10\% false positive rate. We therefore assume that the transiting bodies are indeed planets, with their radii given as in Table~\ref{tab:planetparams}. 

Without direct mass measurements, statistical considerations about the planetary masses can still be performed. \citet{Rogers2015} report that transiting planets with radii above 1.6 Earth radii are typically not rocky, but rather have a gaseous envelope. Our analysis shows that planet b may be rocky, with a radius of $1.31\,R_{\earth}$, while planet d and the candidate with $2.02\,R_{\earth}$ and $1.91\,R_{\earth}$ are likely to have a gaseous envelopes, placing them in the sub-Neptune class. Planet c falls into the intermediate regime with $1.48\,R_{\earth}$, where a rocky and a gas envelope nature are both similarly likely for the planet. 

Using the derived planetary and stellar parameters  and assuming either a rocky or gaseous composition (by using the radius-mass relationship from \citet{fabrycky2014} with values of $\alpha = 3$ and $2.06$ for rocky planets and ones with gas envelopes), the planetary masses would be 2.2, 3.2, 8.2, and 7.0 $M_{\earth}$ for fully rocky planets b, c, d, and the candidate; for gaseous envelopes, the mass estimates amount to 1.7, 2.2, 4.3, and 3.8 $M_{\earth}$.

\footnotetext{Candidate in need of further confirmation.}

\subsection{Prospects for further planet characterisation}

Obtaining precise radial velocity measurements of the host star can be used to determine the planetary masses. Depending on the assumed composition of the planets, we expect radial velocity semi-amplitudes of ca.\ 1\,m/s (gaseous) and up to 3.7 m/s (rocky). This is challenging, but in the feasible regime for near-infrared spectrographs such as CARMENES \citep{carmenes}.

These planets are suitable for atmospheric characterisation with JWST because the star is relatively bright at IR wavelengths. We estimated the amplitude of transit depth variations due to atmospheric absorption by $\Delta D (\lambda) \sim 10 H R_p / R_s^2$, where H is the atmospheric scale height \citep{2010ApJ...716L..74M}. Assuming that a transmission spectrum probes ca.\ five scale heights, with the estimated planetary masses and an Earth-like mean molecular weight, we estimate variations of the order of $10^{-5}$, which are potentially detectable with JWST.

\subsection{Stability of the system}
We tested the dynamic stability of the system using the ``Mercury'' orbital dynamics package \citep{Chambers1999}. We ran the orbital dynamics code with planets b, c and d for 1,000 years (in system time) using the hybrid symplectic/Bulirsch-Stoer integrator assuming initially circular, aligned orbits and the planetary masses from Section~\ref{sec:masses}. The eccentricities of the planets all stay below 0.0003, and the inclination of the orbital planes do not change. We repeated the stability analysis with a replaced planetary mass for the innermost planet in case it is indeed rocky, where we assumed an Earth-like density and found that the results were similar to before. We also ran the code with the planet candidate included and found similar results again; hence we conclude that the system is not strongly dynamically unstable.

\subsection{Habitable zone and evolution considerations}

Habitable zone models by \citet{Kopparapu2013} report the habitable zone for early M dwarfs to range from incident flux levels of ca.\ 0.9 to 0.25 of the flux received by Earth; a more optimistic habitable zone calculation by \citet{Kane2016} puts the inner edge of the habitable zone to flux levels of ca.\ 1.5. Planets b, c, and d are too close to the host star to be habitable. The unconfirmed candidate, however, receives ca.\ 1.9 times the Earth's incident flux if the host star has solar metallicity, and if the host star is metal-poor, this estimate reduces to 1.3, which puts it potentially into the habitable zone.

When the masses of the planets have been determined, comparisons of the planetary densities in the system can give insights into the atmospheric evaporation history of the planets. This has been used successfully by \citet{Lopez2013} for the Kepler-36 system \citep{Carter2012} to show that the planets very likely have different core masses. Similar considerations will be possible for the LP~358-499 system.

\section{Conclusions}
We have presented an analysis of Kepler-K2 light curves of the star LP~358-499. We have determined the star to be an early M dwarf from photometric archival observations, and estimated the distance to the star to be ca.\ 80~pc. We have identified three transiting planets and one candidate planet in the system, with orbital periods of ca.\ 3, 4.9, 11 and 26.6 days and transit depths of ca.\ 700, 1000, 2000 and 1000 ppm. Given the properties of the host star, the smallest of the planets may be rocky. All three planets are closer to the host star than the inner edge of the habitable zone in that system; the planet candidate may be at the inner edge of the habitable zone, depending on stellar parameters. We note that this planetary system is located close to the ecliptic plane and is in fact located in the transit zone of Mercury \citep{Wells2017arXiv}, meaning that transits of Mercury are observable from the location of the LP~358-499 system.

\section*{Acknowledgements}
We thank Guillem Anglada-Escud\'e for helpful comments to improve the manuscript and Martti Kristiansen for aiding in identifying a fourth planet candidate in the system. We also thank Neale Gibson for helpful discussion on atmospheric follow-up possibilities. R.W.\ acknowledges funding from the Northern Ireland Department for Education. K.P.\ and C.W.\ acknowledge funding from the UK Science and Technology Facilities Council.

%%%%%%%%%%%%%%%%%%%%%%%%%%%%%%%%%%%%%%%%%%%%%%%%%%

\bibliographystyle{mnras}
\bibliography{bib}

%%%%%%%%%%%%%%%%%%%%%%%%%%%%%%%%%%%%%%%%%%%%%%%%%%

%%%%%%%%%%%%%%%%% APPENDICES %%%%%%%%%%%%%%%%%%%%%

%%%%%%%%%%%%%%%%%%%%%%%%%%%%%%%%%%%%%%%%%%%%%%%%%%

% Don't change these lines
\bsp	% typesetting comment
\label{lastpage}
\end{document}